\documentclass{mem}
\usepackage{natbib}\usepackage{txfonts}\usepackage{balance}
\usepackage{graphicx}
\usepackage[a4paper]{hyperref}
\idline{75}{282}
\begin{document}
\def\teff{$T\rm_{eff }$}
\def\kms{$\mathrm {km s}^{-1}$}

\title{CODEX/ESPRESSO: the era of precision spectroscopy}

   \subtitle{}

\author{
V. \,D'Odorico on behalf of the CODEX/ESPRESSO team }

  \offprints{V. D'Odorico}

\institute{
Istituto Nazionale di Astrofisica --
Osservatorio Astronomico di Trieste, Via Tiepolo 11,
I-34143 Trieste, Italy
\email{dodorico@oats.inaf.it}
}

\authorrunning{V. D'Odorico}

\titlerunning{CODEX/ESPRESSO: the era of precision spectroscopy}

\abstract{
CODEX is a high resolution spectrograph for the European ELT. CODEX is conceived to reach the highest precision and stability, allowing  the execution of programs spanning many years. Several innovative technical concepts need to be introduced  to reach those excellent characteristics. Thus, the CODEX consortium has foreseen the realization of a CODEX precursor at the VLT: the ESPRESSO spectrograph.  INAF is strongly committed in the ESPRESSO concept study both in terms of financial and human resources. 
\keywords{Cosmology: observations }
}
\maketitle{}

\section{Measuring the expansion of the Universe}

The main goal of the CODEX spectrograph will be the direct measurement of the dynamics of the Universe. It is possible, in principle, to directly  measure the change of the expansion of the Universe with time and compare it with the prediction of the Friedmann-Robertson-Walker ``concordance'' cosmological model. The fundamental importance of this measure relies on its direct nature. Indeed,  current observational constraints on the nature of the ``dark energy'' are basically of geometric kind as they mainly constrain the angular diameter distance to the last scattering surface (Cosmic Microwave Background measurements, e.g. Spergel et al. 2007) and the luminosity distance at moderate redshifts (type Ia Supernovae, e.g. Astier et al. 2006).  

\begin{figure}[]
\resizebox{\hsize}{!}{\includegraphics[clip=true]{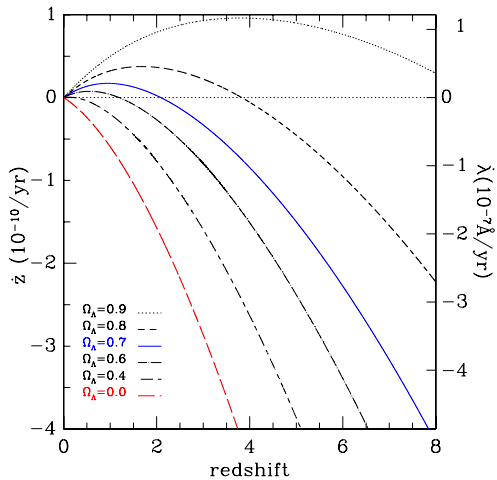}}
\caption{\footnotesize Redshift drift/yr as a function of redshift. The curves refer to relativistic models with no curvature and different values of the cosmological constant (Courtesy of A. Grazian).
}
\label{zpunto}
\end{figure}
The CODEX experiment is conceptually very simple: by making observations of high redshift objects over a time interval of several years, we want to detect and use the wavelength shifts of spectral features of light emitted at high redshift to probe the evolution of the expansion of the Universe directly.  The wavelength change is in fact directly related to the de- or acceleration of the Universe  \citep{grazian}. 
Figure~\ref{zpunto} shows the expected change of redshift for a range of relativistic models with no curvature as a function of redshift. The wavelength shift has a very characteristic redshift dependance. At some redshifts the wavelengths are ``stretched'' while at others they are compressed. The wavelength shift corresponds to a Doppler shift of about 1-10 cm/s over a period of 10 years. 
The numerous absorption lines in the spectra of high-redshift QSOs, which make up the so-called Lyman-$\alpha$ forest, appear to be ideal targets for a measurement of the predicted Doppler shift. 

\begin{figure}[]
\resizebox{\hsize}{!}{\includegraphics[clip=true,angle=270]{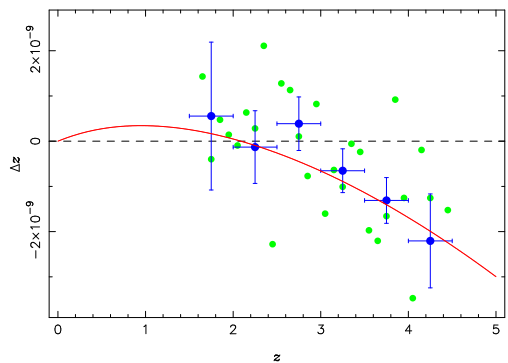}}
\caption{
\footnotesize Full, realistic simulation of the CODEX experiment  for the 
measurement of the cosmic dynamics. Spectra of 36 QSOs at different redshifts 
have been simulated for two different epochs, 20 years apart 
(Courtesy of J. Liske).  
}
\label{sim}
\end{figure}

In order to quantitatively assess the feasibility of the measurement, Monte Carlo simulations have been carried out independently  by several groups \citep{liske}. The high resolution spectra of QSOs were simulated, noise added, and the process repeated for the second epoch. The pairs of spectra   so produced were compared and the measurement performed. Figure~\ref{sim} shows the result of a full simulation, taking 36 QSOs at different redshifts. The cosmological signal is clearly detected.

\section{Immediate science with CODEX}

The scientific application of CODEX, as a high resolution spectrograph with extremely high performance fed by E-ELT, will go beyond the main experiment proposed above. 

Three of the possible outstanding applications are: 

\begin{itemize}
\item[-] {\sl The search for variability of fundamental constants} Fundamental constants are supposedly universal and unvarying quantities. Only astronomical observations hold the the potential to probe the values of fondamental constants in the past and in remote region of space.  In 2001, observations of QSO absorption lines brought the first hints that the value of the fine-structure constant $\alpha$  -- the central parameter in electromagnetism -- might change over time (Murphy et al. 2001), but more recent observations are consistent with a null result. An effective two  to three order of magnitude precision gain is foreseen with a spectrograph with  R~$\approx 150,000$ at E-ELT. The accuracy of the $\Delta \alpha/\alpha$ variation measurements will be  about $10^{-8}$, more precise than any other astronomical and geological measurement.  
 
\item[-] {\sl The search for other earths} Exo-planets and, in particular, terrestrial planets in habitable zones, will be one  of the main scientific topics of the next decades, and one of the main E-ELT science drivers. CODEX with E-ELT will lead the discoveries in at least three main cases in exo-planetary science, providing with unique capabilities and observations: i) discovery and confirmation of rocky planets, ii) search for long-period planets, iii) Jupiter mass planets around faint stars. 
 
\item[-] {\sl Primordial nucleosynthesis} Standard Big Bang nucleosynthesis presents a pressing cosmological conundrum. There is some evidence suggesting a cosmological origin for $^6$Li, and the stellar value for primordial $^7$Li does not agree with primordial Deuterium from QSOs and with $\Omega_{\rm b}$ from WMAP. CODEX will provide the first observations of  $^7$Li and $^6$Li in dwarf stars in galaxies of the Local Group and will make it possible for the first time to measure the interstellar $^7$Li/$^6$Li ratio in unprocessed material of High Velocity Clouds. The latter is a direct and robust probe of the yields of the Big Bang Nucleosynthesis. 
\end{itemize}

\section{ESPRESSO: CODEX precursor at the VLT}

The CODEX concept studies emphasized the need to develop a precursor at the VLT to prove the CODEX  feasibility and in particular to assess some of its development and operational aspects. 
The new instrument for the VLT has been named ESPRESSO for Echelle Spectrograph for PREcision Super  Stable Observations.   ESPRESSO will combine the stability of the HARPS spectrograph at the 3.6m ESO telescope in La Silla with the efficiency of the UVES spectrograph at the VLT. 
In summary, it has to be a high-efficiency, high-resolution, fiber-fed spectrograph of high mechanical and thermal stability and using, if necessary, the simultaneous reference technique. 
It is planned to explore the possibility of using the instrument at the incoherently combined focus of the VLT. This  configuration (equivalent to a 16 m diameter telescope), will push forward the limiting magnitudes of  ESPRESSO and will be useful to test critical components on the way to an equivalent instrument at the E-ELT.    

\subsection{Summary of requirements}

In the standard mode of observation with a single VLT unit, the instrument shall deliver radial-velocity measurements with a precision of 10 cm/sec at any time scale from 20 s up to 30 years. 
However, given some specific scientific goals (Earth-like planets, variability of physical constants, etc.) and the CODEX-precursor nature of ESPRESSO, our goal will be of achieving 1 cm/sec at any time scale from 30 sec up to 30 years.

The definition of the optimal resolving power represents a difficult trade-off because, on the one hand, the highest R should be aimed for, while competition for photons and detector area would call for a low value. The best compromise has been chosen to R~$\sim 150,000$ with a goal for R~$\sim 180,000$ . Good sampling will be assured.

Similarly, we should aim at the largest spectral coverage, compatible with a high system efficiency and affordable design.     The minimum coverage was set to the 370-686 nm range; an extension, both in the red and blue extremes, is foreseen but has to be trade-off with  coatings and fibers length.

 In addition to the CODEX and ESPRESSO projects, we are pursuing an important parallel research: the development   of a novel calibration system based on laser frequency combs \citep{laser}, which will be able to produce  a super accurate, equally spaced, stable source for wavelength calibration. 
 
%\begin{acknowledgements}
%\end{acknowledgements}

\bibliographystyle{aa}

\end{document}